\documentclass[10pt,conference]{IEEEtran}
\IEEEoverridecommandlockouts\IEEEpubid{\makebox[\columnwidth]{ 978-1-6654-3540-6/22~\copyright~2022 IEEE \hfill} \hspace{\columnsep}\makebox[\columnwidth]{ }}
\PassOptionsToPackage{draft}{hyperref}
\usepackage{multirow}
\usepackage{cite}
\usepackage{amsmath,amssymb,amsfonts}
\usepackage{graphicx}
\usepackage{textcomp}
\usepackage{xcolor}
\usepackage{mathtools}
\usepackage{commath}
\usepackage{algorithm}
\usepackage{algpseudocode}
\usepackage{hyperref}
\hypersetup{colorlinks,allcolors=black}
\usepackage[compact]{titlesec}
\titlespacing{\section}{0pt}{2ex}{1ex}
\titlespacing{\subsection}{0pt}{1ex}{0ex}
\titlespacing{\subsubsection}{0pt}{0.5ex}{0ex}
\usepackage{amsmath}

\def\BibTeX{{\rm B\kern-.05em{\sc i\kern-.025em b}\kern-.08em
    T\kern-.1667em\lower.7ex\hbox{E}\kern-.125emX}}
    
\begin{document}
 
\title{Plug-in UL-CSI-Assisted Precoder Upsampling Approach in Cellular FDD Systems}
\author{Yu-Chien Lin, Yan Xin, Ta-Sung Lee, Charlie (Jianzhong) Zhang, Yibo Ma and Zhi Ding
\thanks{Y.-C Lin, Yibo Ma and Z. Ding are with the Department of Electrical and Computer Engineering,
University of California, Davis, CA, USA (e-mail: ycmlin, boma, zding@ucdavis.edu).

Y. Xin and C. Zhang are with Samsung Research America, USA (e-mail: yan.xin, jianzhong.z@samsung.com).

T.-S Lee is with the Institute of Communications Engineering, National Yang Ming Chiao Tung University, Taiwan (e-mail: tslee@nycu.edu.tw).}

\thanks{This work is based on materials supported by the National Science Foundation under Grants 2029027 and 2002937 (Lin, Ding),
by the Samsung GRO Program,  and by the National Science \& Technology Council of Taiwan grants NSTC 112-2218-E-A49-020, NSTC 112-2221-E-A49-079 (Lee).}}

\maketitle
\pagestyle{empty}
\thispagestyle{empty}
\begin{abstract}
Acquiring downlink channel state information (CSI) is crucial for optimizing performance in massive Multiple Input Multiple Output (MIMO) systems operating under Frequency-Division Duplexing (FDD). Most cellular wireless communication systems employ codebook-based precoder designs, which offer advantages such as simpler, more efficient feedback mechanisms and reduced feedback overhead. Common codebook-based approaches include Type II and eType II precoding methods defined in the 3GPP standards. Feedback in these systems is typically standardized per subband (SB), allowing user equipment (UE) to select the optimal precoder from the codebook for each SB, thereby reducing feedback overhead. However, this subband-level feedback resolution may not suffice for frequency-selective channels. This paper addresses this issue by introducing an uplink CSI-assisted precoder upsampling module deployed at the gNodeB. This module upsamples SB-level precoders to resource block (RB)-level precoders, acting as a plug-in compatible with existing gNodeB or base stations.

\end{abstract}

\section{Introduction}
Massive multiple-input multiple-output (MIMO) technology significantly enhances spectrum and energy efficiency in wireless systems. However, it requires precise downlink (DL) channel state information (CSI) acquisition at the base station or gNodeB (gNB). In frequency-division duplexing (FDD) systems, DL CSI acquisition relies on user equipment (UE) feedback, which can be costly due to the large number of channel coefficients. Efficient compressive CSI feedback is crucial to conserve uplink (UL) bandwidth and UE power, enabling the practical deployment of massive MIMO in FDD networks. CSI feedback can be explicit or implicit.\\

In explicit CSI feedback, the DL CSI is compressed on the UE side by exploiting its sparsity in the beam and delay domains and then recovered by the gNB. A prominent explicit framework is the deep autoencoder, as demonstrated in \cite{CsiNet}, which includes an encoder at the UE and a decoder at the gNB. Various autoencoder models, such as those presented in \cite{CsiNet-LSTM, CRNet, DeepCMC, CLNet, SCEnet, ZR, 6G}, have shown superior CSI recovery or lightweight designs. Recent approaches have utilized underlying channel correlation to improve DL CSI recovery at base stations, leveraging previous CSI \cite{CsiNet-LSTM, MarkovNet}, CSI of nearby UEs \cite{CoCsiNet}, and UL CSI \cite{CQNET, DualNet, DualNet-MP}. Considering practical sparse pilot placement, SRCsiNet \cite{SRCsiNet} has been proposed as a DL CSI upsampling module to handle aliasing effects in the delay domain for high delay-spread (DS) DL CSIs.\\

\begin{figure}
    \centering
    \resizebox{3.4in}{!}{
    \includegraphics{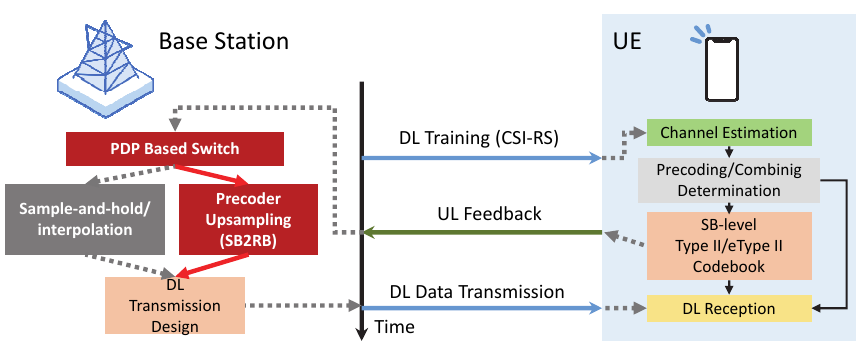}}
    \caption{Illustration of the proposed precoder upsampling approach and the considered FDD system.}
    \label{fig: SRPNet_Architecture}
\end{figure}

In implicit CSI feedback, instead of feeding back the DL CSI, the UE selects a precoder from a codebook based on the estimated DL CSI and sends the precoder matrix indicator (PMI) to the gNB. Current cellular systems adopt codebook-based approaches such as Type II \cite{3GPPtypeII} and eType II \cite{3GPPetypeII}. To reduce UL feedback overhead, precoders are selected and fed back for each subband (SB). However, the aliasing issue becomes more severe due to SB-level precoder feedback. As illustrated in Figure \ref{fig: SRPNet_Architecture}, operators aim to find a precoder upsampler to map SB-level precoders to RB-level ones (i.e., SB2RB), achieving higher DL throughput. However, due to limited SB-level feedback resolution, interpolating missing RB-level precoders is challenging.\\



Our primary objective is to address the undersampling issue caused by SB-level precoder feedback in Type II/eType II feedback of the existing cellular network standards. We introduce an SB2RB precoder upsampling methodology utilizing UL CSI to design a bandpass filter that mitigates the undersampling problem. We develop a physics-inspired deep learning architecture that leverages UL CSI for effective aliasing suppression and acts as a plug-in module compatible with current codebook-based approaches. Additionally, we design a switch to determine whether to apply or bypass the proposed network to reduce complexity. Our key contributions are as follows:

\begin{itemize}
    \item We establish a deep learning framework, SRPNet, consisting of three modules: precoder initial upsampling, UL-CSI-assisted bandpass-filter design, and upsampling refinement modules. This framework effectively mitigates aliasing due to SB-level feedback.
    
    \item Compared to interpolation, SRPNet has higher complexity. For most precoders of low-DS DL CSIs, interpolation suffices to recover RB-level precoders from SB-level ones. Thus, a PDP-based switch is designed to decide when to apply SRPNet for complexity reduction.

\end{itemize}
\section{System Model}
\subsection{Downlink Transmission in FDD System}
In the system setup, a gNB equipped with $N_a$ antennas establishes communication with a single-antenna UE. The downlink transmission employs a precoding technique to optimize the signal's transmission over the wireless channel. Considering the $f$-th RB, the signal received on the UE side can be expressed as a linear combination of the transmitted symbols and noise. Mathematically, the received signal $y_f$ can be represented as:
\begin{gather*}
    y_f = \mathbf{h}_f^H\mathbf{w}_fs_f + n_f.
\end{gather*}
Where $\mathbf{h}_f \in \mathbb{C}^{N_a \times 1}  $ represents the channel matrix corresponding to the $f$-th RB, $\mathbf{w}_f$ denotes the precoding vector applied by the gNB, $s_f$ is the symbol transmitted over the f-th RB, and $n_f$ accounts for the additive noise. The precoding vector $\mathbf{w}_f$ optimizes signal transmission by matching the channel.

\subsection{Codebook Based Precoder Design}
The idea of codebook-based precoder is to determine the precoder at UE side from codebooks and feedback to gNB for the following downlink transmission. There are shared codebooks between gNB and UEs. UEs estimate DL CSI from pilots, and then feedback precoder matrix indicator (PMI) and layer indicator (LI) to gNB. Common codebook-based precoding in modern FDD system include Type I, Type II and eType II \cite{3GPPtypeI, 3GPPtypeII, 3GPPetypeII} precoding, which will be introduced below in a high-level manner (for simplicity, we consider one polarization only):

\subsubsection{Type I/Type II Precoding}
Type I precoding exploits the spatial diversity provided by multiple transmit antennas to enhance communication performance. The UE selects a beam and a co-phase coefficient from an oversampled set of beam directions as the Type I precoder. The Type I precoder for the $f$-th SB can be expressed as:

\begin{equation}
    \mathbf{w}_f = \text{argmax}_{\mathbf{w}\in\Omega_l}\{|\mathbf{h}_f^H\mathbf{w}|\}, f = 1,...,N_3.
    \label{TypeI precoder}
\end{equation}
Where $\Omega_l$ is the codebook containing oversampled beams corresponding to the $l$-th layer, and $N_3$ is the number of SBs in BWP.
Then the UE acknowledges the selected precoder by feeding back the beam index (i.e., PMI) and the layer $l$. For the Type II precoder, designed for the multi-user MIMO (MU-MIMO) use case, it provides more flexibility in choosing multiple beams and the degree of freedom to combine the selected beams to match DL CSI. The selected Type II precoder for the $f$-th SB can be expressed as:

\begin{gather*}
    \mathbf{w}_f = \sum_{i = 1}^{L}{\alpha_{f,i}\mathbf{w}_{f,i}/L}.
    \label{TypeII precoder}
\end{gather*}
where $\mathbf{w}_{f,i}$ is the $i$-th selected oversampled beam, and $\alpha_{f,i}$ is the complex combining coefficient in the $f$-th SB. Note that, to reduce the feedback overhead, both Type I and Type II precoders are fed back per SB.

\subsubsection{eType II Precoding}
\begin{figure*}
    \centering
    \resizebox{7in}{!}{
    \includegraphics{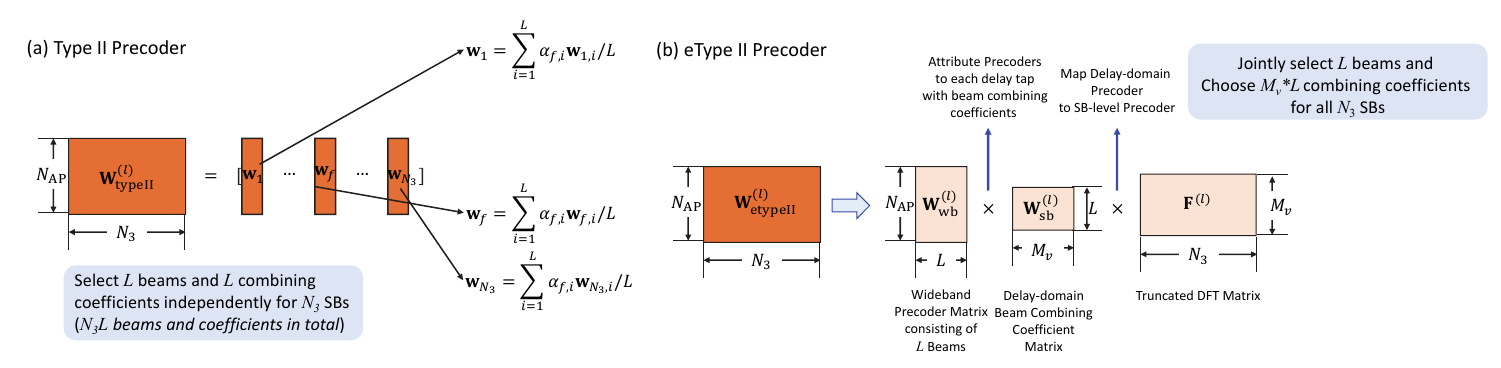}}
    \caption{The comparison between (a) Type II and (b) eType II Precoder.}
    \label{fig: Type II vs eType II}
\end{figure*}

The eType II precoder is a more efficient feedback method compared to Type I and Type II precoders. Figure \ref{fig: Type II vs eType II} reveals the differences between Type II and eType II precoding. In Figure \ref{fig: Type II vs eType II}(a), it can be seen that the UE designs the Type II precoder for each SB independently. Considering the high correlation of the spatial structures of channels for SBs in a BWP, the eType II precoder design allows the gNB to enable UEs to jointly select $L$ wideband beams and feedback the precoder for all SBs in the entire BWP, a process called spatial compression.

To further reduce the feedback overhead, the eType II precoder performs frequency-domain compression. It first transforms the SB-level precoders into delay-domain ones. According to the principles of radiology and propagation loss, the delay-domain beam combining coefficients are truncated by retaining only the first $M_v$ delay taps. To further compress in the delay domain, due to the sparsity of the truncated delay-domain coefficients, the coefficients can be compressed by a factor of $R$, feeding back only the significant delay taps and their positions shown in the left part of Figure \ref{fig: ModifedEtypeII}.

\subsection{Problem Formulation}
Given the restriction of uplink feedback overhead, all types of codebook-based precoder feedback can only be conducted per SB. For some frequency-selective channels (e.g., outdoor channels with large delay spread), such a low feedback rate cannot fully exploit the channel diversity provided by the precoder. Without modifying the current specification, operators seek a non-linear mapping function $f_\Theta(\cdot)$ to upsample the SB-level precoders to RB-level ones, thereby better exploiting the channel gain via a finer-resolution precoder in the frequency domain. The loss function can be expressed as follows:

\begin{equation}
    \Theta = \text{argmax}_\Theta\sum_{f=0}^{N_\text{RB}}|\mathbf{h}_f^Hf_\Theta(\mathbf{w}_{\left \lfloor{f/N_\text{RBpSB}}\right \rfloor})|.
\end{equation}
where $\Theta$ represents the trainable parameters of the learning-based upsampler, $\mathbf{w}_f$ is the precoder for the $f$-th SB, and ${\left\lfloor f/N_\text{RBpSB} \right\rfloor}$ is a floor operation to transform from RB index to SB index. $N_\text{RBpSB}$ is a system parameter that determines the number of RBs in an SB.

\section{Type II/eTypeII based Precoder Upsampling}

\subsection{General Architecture}
We propose a lightweight network deployed at the gNB that acts as a plug-in module, providing precoder upsampling from SB-level to RB-level. This architecture is compatible with existing modern cellular systems, such as 5G-NR. Figure \ref{fig: SRPNetModel} provides a high-level illustration of the proposed architecture, SRPNet. This network can effectively recover undersampled channels by exploiting the DFT shifting invariance property. Due to the UL/DL path reciprocity, the network can significantly suppress the aliasing effects caused by sub-Nyquist sampling. The details can be found in \cite{SRCsiNet}.

\subsection{Modified Type II/eType II precoding}
We discovered that the selected precoders according to Eq. \ref{TypeI precoder} may lose multipath delay information. To maintain the signal structure of DL CSIs in Type II/eType II precoder design, we modify the precoder design criterion as follows:

\begin{equation}
    \mathbf{w}_f = \text{argmin}_\mathbf{w}\norm{\frac{\mathbf{h}_f}{\norm{\mathbf{h}_f}_2} - \mathbf{w}}_2 \label{eq: new criterion}
\end{equation} 
This criterion ensures that the Type II/eType II precoder is close to the normalized DL CSI and preserves the phase information. Note that this criterion does not apply to the Type I precoder, as it lacks the degrees of freedom in choosing beams and combining coefficients.

\begin{figure}
    \centering
    \resizebox{3.4in}{!}{
    \includegraphics{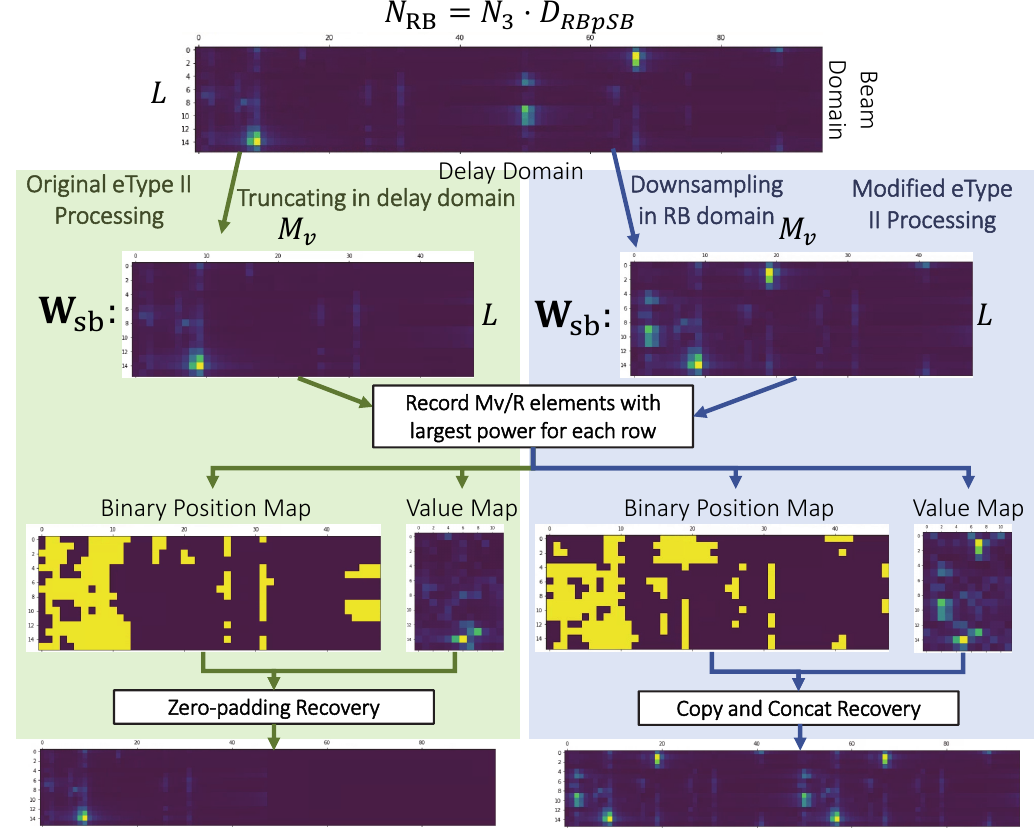}}
    \caption{Illustration of the original and the modified eType II encoding processing for precoders of a high-DS DL CSI.}
    \label{fig: ModifedEtypeII}
\end{figure}

For the eType II precoder, there is a critical problem to be solved. In the eType II precoder, to reduce the feedback overhead, as illustrated in the left part of Figure \ref{fig: ModifedEtypeII}, truncation is performed, leaving the remaining delay components zero except for the first $M_v$ delay taps. This truncation in the delay domain seems reasonable for most low-delay spread (DS) channels but performs poorly in capturing the high-delay components for precoders of high DS channels. Once the truncation is done on the UE side, it is impossible to recover on the gNB side. In the proposed approach, we replace the delay-domain truncation with frequency-domain downsampling. As illustrated in the right part of Figure \ref{fig: ModifedEtypeII}, we uniformly sample $M_v$ precoders in the RB domain and then transform them into the delay domain. We find that this modification preserves all the multipath information but might mistake low-delay components for high-delay ones due to sub-Nyquist sampling in the frequency domain, leading to aliasing effects. However, these aliasing effects can be alleviated by the following precoder upsampling module, SRPNet.

\subsection{Precoder Upsampler, SRPNet}
With the aid of delay domain sparsity, we introduce a lightweight neural network (NN), called the super-resolution precoder network (SRPNet), to suppress the aliasing effect due to sub-Nyquist sampling.

\subsubsection{Architecture}
\begin{figure}
    \centering
    \resizebox{3.4in}{!}{
    \includegraphics{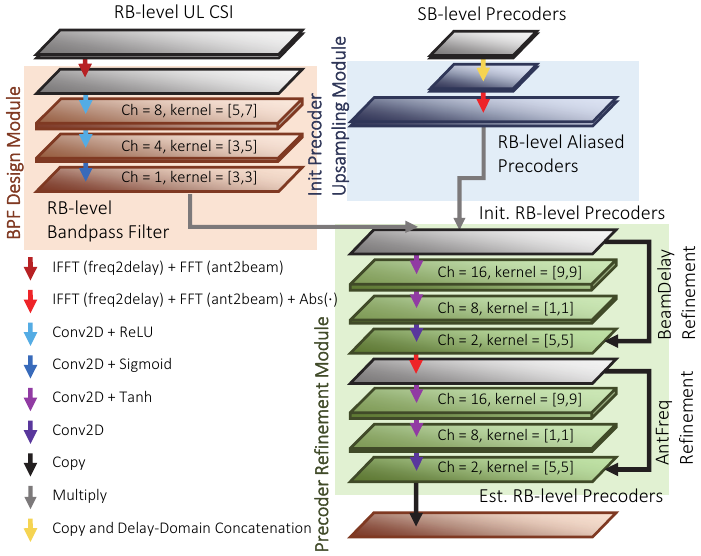}}
    \caption{The model design of the SRPNet. It consits of three modules: 1) BPF Design Module, 2) Initial Precoder Upsampling Module, and 3) Precoder Refinement Module.}
    \label{fig: SRPNetModel}
\end{figure}

This network consists of three modules: 1) bandpass filter (BPF) Design Module, 2) Initial Precoder Upsampling Module, and 3) Precoder Refinement Module, as illustrated in Figure \ref{fig: SRPNetModel}.

\begin{itemize}
    \item \textbf{BPF Design Module}: To obtain the delay profile of precoders, leveraging UL/DL multipath reciprocity, we feed the delay profile of the UL CSI into a convolutional network to infer a BPF that suppresses aliasing delay taps and preserves the true delay peaks of initial RB-level precoders in the delay domain.
    
    \item \textbf{Initial Precoder Upsampling Module}: We feed the modified Type II/eType II SB-level precoders to generate RB-level aliased precoders as initial precoders. These precoders may suffer severe aliasing effects but preserve all the true delay peaks at the same time.

    \item \textbf{Precoder Refinement Module}: First, we perform element-wise multiplication of the BPF and the RB-level aliased precoders. This guides the BPF Design Module to design a BPF instead of a confusing matrix\footnote{If the BPF Design Module does not give a BPF, the aliasing peaks that are not suppressed will confuse the rest of the network.}. This opens up part of the black box of the NN. Then, we perform BD-domain and AF-domain refinement using convolutional NNs and shortcuts to generate the estimated RB-level precoders.

\end{itemize}

Lastly, we design the network with fully convolutional layers for scalability to any size of the input (i.e., different array sizes and bandwidths). For practical deployment considerations, SRPNet may not always be necessary for different scenarios such as low-DS DL CSIs. In the next section, we provide a PDP-based switch to determine when to utilize SRPNet.

\section{UL-CSI/SSB Assisted Switch for Low-Complexity Precoder Upsampling}
This section aims to provide a bridge for the proposed approach to practical cellular systems. We describe a PDP-based switch to help the system decide when to use simple linear interpolation or SRPNet to upsample SB-level precoders to RB-level.

Even with the design of an extremely lightweight and flexible network for precoder upsampling, Table \ref{tab:complexity_analysis} shows that SRPNet still leads to much higher computational complexity compared to linear interpolation. Given that for most channels with low delay spread (DS), linear interpolation can provide good enough precoder upsampling. Therefore, in this section, we propose several options to switch between SRPNet and linear interpolation.

\subsection{PDP-based Switch}
The key to determining whether to use SRPNet lies in evaluating how frequently the CSI varies with different RBs, which can be inferred from its delay profile. Although the gNB does not have the power delay profile (PDP) of DL CSI, it can still infer it from the PDP correlation between UL and DL CSIs.

\subsubsection{Threshold-based Switch}
Assume we have RB-level PDP from UL CSI or SSB $\mathbf{PDP} \in \mathbb{C}^{N_\text{RB}}$. We make a decision $s$ (i.e., $s=1$ means utilization of SRPNet and vice versa) by applying the following measures, $m$, to trigger the switch on or off:

\begin{gather*}
s = \left\{
\begin{aligned}
    1, & \ \ m \geq thres,\\
    0, & \ \ \text{otherwise.}
\end{aligned}
\right.
\end{gather*}
\begin{itemize}
    \item Maximum Excess Delay: the delay difference between significant multipath components.
    \item Mean Excess Delay: the mean delay weighted by its PDP. 
    \item Root-Mean-Square (RMS) DS 
\end{itemize}

\subsubsection{Learning-based Switch}
We also train a learning-based switch with a single-layer NN, which can be represented as:

\begin{gather*}
    s = f_\text{switch}(\mathbf{PDP}) = Sigmoid(\mathbf{f}^{T}\mathbf{PDP}+b),
\end{gather*}
by maximizing the gain-minus-cost metric $G - \lambda \cdot C$, where $G = s \cdot NG(\mathbf{W}_\text{SRPNet}, \mathbf{H}) + (1-s) \cdot NG(\mathbf{W}_\text{ITP}, \mathbf{H})$ and $C = s \cdot C_\text{SRPNet} + (1-s) \cdot C_\text{ITP}$. Here, $NG(\cdot)$ is the function to evaluate the average normalized gain between the precoders and DL CSIs. $C_\text{SRPNet}$ and $C_\text{ITP}$ are the computational complexities of SRPNet and linear interpolation, respectively. The ratio $\frac{C_\text{SRPNet}}{C_\text{ITP}}$ is roughly 1000. $\lambda$ is a hyperparameter that determines the weight between computational cost and performance. In the test stage, we round $s$ to determine the final outcome.

\section{Experimental Evaluations}
\subsection{Experiment Setup}
Tests were focused on outdoor channels using the widely used channel model software, QuaDriGa. The simulator considers a gNB with a 128-element uniform linear array (ULA) serving single-antenna UEs, with half-wavelength uniform spacing. 2000 UEs are uniformly distributed in the cell coverage, which is a rectangular region of size $250\,\text{m} \times 300\,\text{m}$. The scenario features given in 3GPP TR 38.901 UMa were followed, using $N_\text{RB} = 96$ resource blocks (RBs) with a $20$ MHz bandwidth part. The normalized gain
\begin{gather*}
g = \sum_{f=1}^{N_\text{RB}} \frac{|\mathbf{h}_f^H \mathbf{w}|}{|\mathbf{h}_f| \cdot |\mathbf{w}|}
\end{gather*}
was used to assess performance.

For DL-based models, we conducted training with a batch size of 32 for 1500 epochs, starting with a learning rate of 0.001 and setting an early stop criterion if the validation loss did not improve for 100 epochs. We generated the outdoor datasets using the QuaDRiGa channel simulators. We considered 16 transmission time intervals (TTIs) for each of the 2000 UEs. In total, the dataset consists of 32,000 channels. We used one-tenth of the channels for testing and validation, respectively. The remaining four-fifths of the channels were used for training.

To evaluate the degree of aliasing, it is common to use DS as a performance metric. A channel with a larger DS tends to suffer from aliasing effects more severely since it contains more high-delay multipaths. We clustered all the 3200 test CSI data into 3 clusters according to their RMS DS: low (smaller than 500 ns), medium (between 500 ns and 1000 ns), and high DS (larger than 1000 ns). The low, medium, and high DS clusters have 883, 1221, and 1095 test cases, respectively.

\subsection{Applying SRPNet to SB-level Type II Precoder}

\begin{figure}
    \centering
    \resizebox{3in}{!}{
    \includegraphics{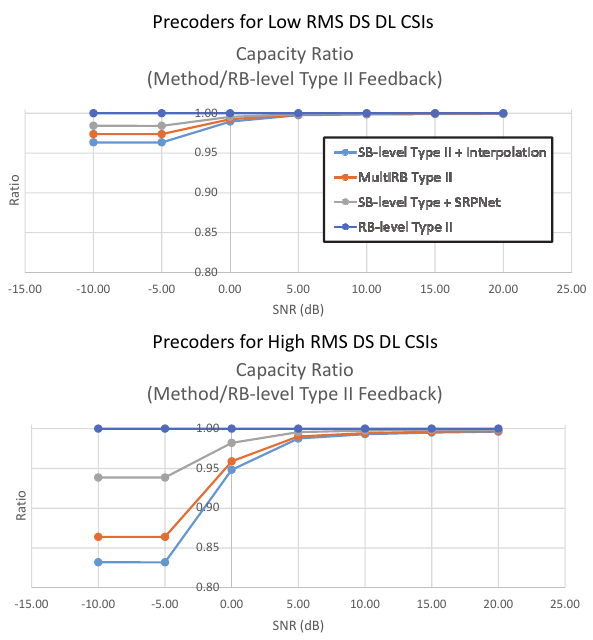}}
    \caption{Capacity ratio of the SRPNet and other codebook-based approaches. Upper one is the results for the low RMS DS cluster. Bottom one is for the large DS cluster)}
    \label{fig: Sim_TypeIIvsSRPNet}
\end{figure}

Figure \ref{fig: Sim_TypeIIvsSRPNet} shows the capacity improvement ratio of precoders after applying SRPNet at different SNRs for low and high DS CSIs. Apparently, SRPNet improves capacity significantly especially for low SNRs. In addition, we can find that the benefit of the SRPNet becomes more obvious in high DS CSIs. That is because aliasing effect occurs for high DS CSIs with higher probability and SRPNet can effectively upsample SB-level precoders to RB-level ones even if aliasing effects exist.

\subsection{Applying SRPNet to SB-level eType II Precoder}

\begin{figure}
    \centering
    \resizebox{3in}{!}{
    \includegraphics{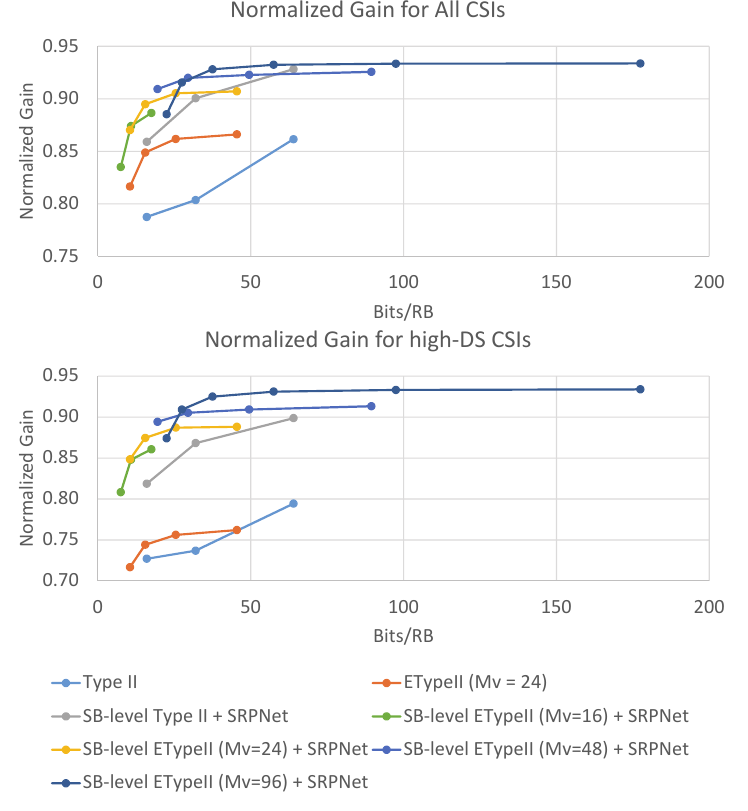}}
    \caption{Normalized gain of Type II and eType II precoders before and after being applied SRPNet for all DL CSIs and high DS ones.}
    \label{fig: Sim_EtypeIIvsSRPNet}
\end{figure}

Figure \ref{fig: Sim_EtypeIIvsSRPNet} demonstrates the normalized gain of Type II and eType II precoders before and after being applied SRPNet with different settings. Different points in a curve represent different configurations. For Type II precoders, we consider four different numbers of SBs ($N_3 = 3, 6, 12, 24$) representing different frequency downsampling rates (from 96 RBs). For each curve of the eType II precoder, the anchor points from right to left represent $R = 2, 4, 8, 16$, and so on.

We observe a significant performance gap between the eType II precoder ($M_v = 24$) with and without SRPNet upsampling for high DS scenarios. Additionally, SRPNet-based eType II precoders outperform Type II and SRPNet-based Type II precoders, especially for low UL feedback overhead. This demonstrates the higher efficiency of the eType II precoder compared to Type II precoders after applying SRPNet. However, we also find that eType II precoders do not perform better even with $R = 1$. This is because eType II precoders find a common set of $L$ beams for all SBs, which may not be optimal for each SB, leading to a performance bound.

\subsection{Applying PDP-based Switch for Complexity Reduction}
\begin{figure}
    \centering
    \resizebox{3.4in}{!}{
    \includegraphics{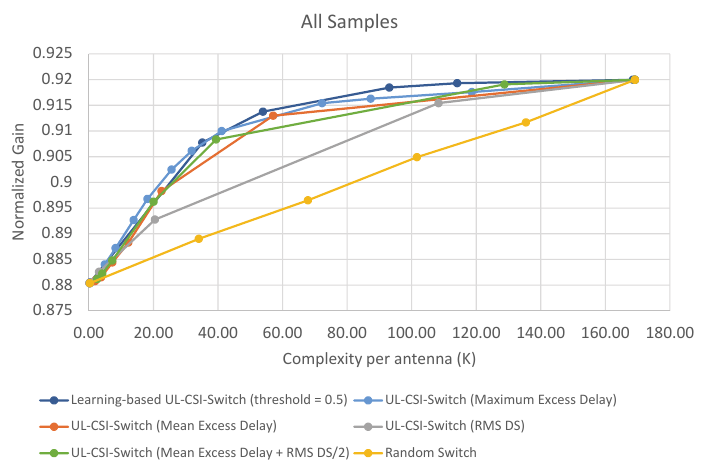}}
    \caption{Normalized gain after applying the proposed PDP-based switches and random switch for computational complexity reduction of precoder upsampling.}
    \label{fig: Sim_Learning-ULCSI-Switch}
\end{figure}

Figure \ref{fig: Sim_Learning-ULCSI-Switch} shows the normalized gain after applying the proposed PDP-based switches and a random switch for computational complexity reduction of precoder upsampling. We compare our proposed switches (Threshold-based Switches and Learning-based Switch) with a baseline random switch, which randomly chooses to utilize SRPNet or interpolation. The curve of the random switch is generated by setting different probabilities $p$ to choose SRPNet ($p = 1$ for the rightmost point). It forms a straight line since both the normalized gain and complexity are linear combinations of the outcomes of SRPNet and interpolation. We find that all the proposed switches perform better than the random switch, indicating that the PDP of UL CSI or SSB is beneficial for making the binary decision.

Among these rule-based switches, the one relying on maximum excess delay performs the best since maximum excess delay is more direct and can better reflect when aliasing occurs (i.e., when the largest delay of significant paths exceeds the Nyquist measurable delay). The curve of the learning-based switch is built by training the model with different $\lambda = 1 \times 10^{-5}, 5 \times 10^{-5}, \ldots, 1 \times 10^{-3}$. The learning-based approach performs the best among all the proposed switches and also has the lowest complexity. However, we still face the challenge of finding a mapping from choosing $\lambda$ to achieving the desired complexity.

\section{Conclusions}
Acquiring accurate DL CSI is crucial for optimizing the performance of massive MIMO systems in FDD. Existing cellular systems use codebook-based precoder designs, such as Type II and eType II, which simplify feedback mechanisms and reduce overhead. However, feedback standardized per SB often falls short for frequency-selective channels.
To address this issue, we introduced SRPNet, an uplink CSI-assisted precoder upsampling module deployed at the gNodeB. SRPNet enhances SB-level precoders to RB-level precoders and is compatible with existing base stations. Our results demonstrated SRPNet's effectiveness in improving normalized gain, particularly in high DS scenarios. Additionally, we proposed a PDP-based switch to intelligently choose between SRPNet and linear interpolation, reducing computational complexity. Our findings showed that the proposed switches, especially the learning-based switch, outperformed random switches and achieved better performance with lower complexity. In summary, SRPNet and the PDP-based switch offer a robust solution for enhancing downlink CSI acquisition in massive MIMO systems. These advancements significantly improve the efficiency and performance of modern cellular networks, particularly in scenarios with high-frequency selectivity.

\bibliography{references.bib}

\begin{thebibliography}{10}
\providecommand{\url}[1]{#1}
\csname url@samestyle\endcsname
\providecommand{\newblock}{\relax}
\providecommand{\bibinfo}[2]{#2}
\providecommand{\BIBentrySTDinterwordspacing}{\spaceskip=0pt\relax}
\providecommand{\BIBentryALTinterwordstretchfactor}{4}
\providecommand{\BIBentryALTinterwordspacing}{\spaceskip=\fontdimen2\font plus
\BIBentryALTinterwordstretchfactor\fontdimen3\font minus \fontdimen4\font\relax}
\providecommand{\BIBforeignlanguage}[2]{{%
\expandafter\ifx\csname l@#1\endcsname\relax
\typeout{** WARNING: IEEEtran.bst: No hyphenation pattern has been}%
\typeout{** loaded for the language `#1'. Using the pattern for}%
\typeout{** the default language instead.}%
\else
\language=\csname l@#1\endcsname
\fi
#2}}
\providecommand{\BIBdecl}{\relax}
\BIBdecl

\bibitem{CsiNet}
C.~{Wen}, W.~{Shih}, and S.~{Jin}, ``{D}eep {L}earning for {M}assive {MIMO} {CSI} {F}eedback,'' \emph{IEEE Wirel. Commun. Lett.}, vol.~7, no.~5, pp. 748--751, 2018.

\bibitem{CsiNet-LSTM}
T.~Wang, C.-K. Wen, S.~Jin, and G.~Y. Li, ``Deep learning-based csi feedback approach for time-varying massive mimo channels,'' \emph{IEEE Wireless Communications Letters}, vol.~8, no.~2, pp. 416--419, 2019.

\bibitem{CRNet}
Z.~Lu, J.~Wang, and J.~Song, ``{Multi-resolution CSI Feedback with Deep Learning in Massive MIMO System},'' in \emph{IEEE Intern. Conf. Communications (ICC)}, 2020, pp. 1--6.

\bibitem{DeepCMC}
Q.~Yang, M.~B. Mashhadi, and D.~Gündüz, ``{Deep Convolutional Compression For Massive MIMO CSI Feedback},'' in \emph{IEEE Intern. Workshop Mach. Learning for Signal Process. (MLSP)}, 2019, pp. 1--6.

\bibitem{CLNet}
S.~Ji and M.~Li, ``{CLNet: Complex Input Lightweight Neural Network Designed for Massive MIMO CSI Feedback},'' \emph{IEEE Wirel. Commun. Lett.}, vol.~10, no.~10, pp. 2318--2322, 2021.

\bibitem{SCEnet}
Y.-C. Lin, T.-S. Lee, and Z.~Ding, ``{A Scalable Deep Learning Framework for Dynamic CSI Feedback with Variable Antenna Port Numbers},'' \emph{IEEE Trans. Wirel. Commun.}, vol.~23, no.~4, pp. 3102--3116, 2024.

\bibitem{ZR}
------, ``{Training-Free Cost-Efficient Compression for Massive MIMO Channel State Feedback},'' in \emph{GLOBECOM 2023}, 2023, pp. 3391--3396.

\bibitem{6G}
C.-H. Lin, S.-C. Lin, and L.~C. Chu, ``{A Low-Overhead Dynamic Formation Method for LEO Satellite Swarm Using Imperfect CSI},'' \emph{IEEE Trans. Veh. Tech.}, vol.~73, no.~5, pp. 6923--6936, 2024.

\bibitem{MarkovNet}
Z.~{Liu}, M.~{Rosario}, and Z.~{Ding}, ``{A} {M}arkovian {M}odel-{D}riven {D}eep {L}earning {F}ramework for {M}assive {MIMO} {CSI} {F}eedback,'' \emph{IEEE Trans. Wirel. Commun.}, vol.~21, no.~2, pp. 1214--1228, 2022.

\bibitem{CoCsiNet}
J.~{Guo \textit{et al.}}, ``{DL}-based {CSI} {F}eedback and {C}ooperative {R}ecovery in {M}assive {MIMO},'' \emph{arXiv preprint arXiv:2003.03303}, 2020.

\bibitem{CQNET}
Z.~{Liu}, L.~{Zhang}, and Z.~{Ding}, ``{A}n {E}fficient {D}eep {L}earning {F}ramework for {L}ow {R}ate {M}assive {MIMO} {CSI} {R}eporting,'' \emph{IEEE Trans. Commun.}, vol.~68, no.~8, pp. 4761--4772, 2020.

\bibitem{DualNet}
------, ``{E}xploiting {B}i-{D}irectional {C}hannel {R}eciprocity in {D}eep {L}earning for {L}ow {R}ate {M}assive {MIMO} {CSI} {F}eedback,'' \emph{IEEE Wirel. Commun. Lett.}, vol.~8, no.~3, pp. 889--892, 2019.

\bibitem{DualNet-MP}
Y.-C. Lin, Z.~Liu, T.-S. Lee, and Z.~Ding, ``{D}eep {L}earning {P}hase {C}ompression for {MIMO} {CSI} {F}eedback by {E}xploiting {FDD} {C}hannel {R}eciprocity,'' \emph{IEEE Wireless Commun. Lett.}, vol.~10, no.~10, pp. 2200--2204, 2021.

\bibitem{SRCsiNet}
Y.-C. Lin, Y.~Xin, T.-S. Lee, C.~Zhang, and Z.~Ding, ``{Physics-Inspired Deep Learning Anti-Aliasing Framework in Efficient Channel State Feedback},'' \emph{arXiv preprint arXiv:2403.08133}, 2024.

\bibitem{3GPPtypeII}
3GPP, ``Type {II} {CSI} {R}eporting,'' TSG RAN WG1 No. 89 R1-1707962, 2021.

\bibitem{3GPPetypeII}
------, ``{Physical layer procedures for data },'' 3GPP TS 38.214 version 16.2.0 Release 16, 2020.

\bibitem{3GPPtypeI}
------, ``On {NR} {T}ype {I} {C}odebook,'' TSG RAN WG1 No. 88 R1-1702205, 2021.

\end{thebibliography}
\bibliographystyle{IEEEtran}
\end{document}